\documentclass[english,prd,english,notitlepage,showpacs,preprintnumbers,showkeys,nofootinbib]{revtex4}
\usepackage[latin9]{inputenc}
\usepackage{babel}
\usepackage{amsmath}
\usepackage{amssymb}
\usepackage{graphicx}
\usepackage{esint}
\usepackage[unicode=true,pdfusetitle,
 bookmarks=true,bookmarksnumbered=false,bookmarksopen=false,
 breaklinks=false,pdfborder={0 0 1},backref=false,colorlinks=false]
 {hyperref}
\usepackage{breakurl}

\makeatletter

\providecommand{\tabularnewline}{\\}

\@ifundefined{textcolor}{}
{%
 \definecolor{BLACK}{gray}{0}
 \definecolor{WHITE}{gray}{1}
 \definecolor{RED}{rgb}{1,0,0}
 \definecolor{GREEN}{rgb}{0,1,0}
 \definecolor{BLUE}{rgb}{0,0,1}
 \definecolor{CYAN}{cmyk}{1,0,0,0}
 \definecolor{MAGENTA}{cmyk}{0,1,0,0}
 \definecolor{YELLOW}{cmyk}{0,0,1,0}
}

\usepackage{babel}

\makeatother

\begin{document}

\title{Higher-twist contributions to neutrino-production of pions}

\author{B.~Z.~Kopeliovich, Iván~Schmidt and M.~Siddikov}

\address{Departamento de F\'isica, Universidad T\'ecnica Federico Santa Mar\'ia,\\
 y Centro Cient\'ifico - Tecnológico de Valpara\'iso, Avda. Espa\~na 1680, Valpara\'iso,
Chile}

\preprint{USM-TH-318}
\begin{abstract}
In this paper we estimate the size of twist-3 corrections to the deeply
virtual meson production in neutrino interactions due to the chiral odd
transversity Generalized Parton Distribution (GPD). We conclude
that in contrast to pion \emph{electro}production, in neutrino-induced reactions
these corrections are small. This happens due to large contribution
of unpolarized GPDs $H,\, E$ to the leading-twist amplitude in neutrinoproduction.
We provide a computational code, which can be used for evaluation
of the cross-sections accounting for these twist-3 corrections with
various GPD models. Our results are particularly relevant for analyses of the pion
and kaon production in the \textsc{Minerva }experiment at FERMILAB. 
\end{abstract}



\maketitle

\section{Introduction}

Nowadays one of the key objects used to parametrize nonperturbative
structure of the target are the generalized parton distributions (GPDs).
For the kinematics where the collinear factorization is applicable~\cite{Ji:1998xh,Collins:1998be},
they allow evaluation of the cross-sections for a wide class of processes.
Today all information on GPDs comes from the electron-proton and
positron-proton measurements performed at JLAB and HERA, in particular
from deeply virtual Compton scattering (DVCS) and deeply virtual meson
production (DVMP)~\cite{Ji:1998xh,Collins:1998be,Mueller:1998fv,Ji:1996nm,Ji:1998pc,Radyushkin:1996nd,Radyushkin:1997ki,Radyushkin:2000uy,Collins:1996fb,Brodsky:1994kf,Goeke:2001tz,Diehl:2000xz,Belitsky:2001ns,Diehl:2003ny,Belitsky:2005qn,Kubarovsky:2011zz}.
The forthcoming CLAS12 upgrade at JLAB will help to improve our understanding
of the GPDs~\cite{Kubarovsky:2011zz}. However, in practice, the procedure of extraction
of GPDs from experimental data is subject to the uncertainties,
like large BFKL-type logarithms in the next-to-leading order (NLO) corrections~\cite{Ivanov:2007je}
in the HERA kinematics; contributions of the higher-twist components of GPDs and the pion distribution
amplitudes (DAs) in the JLAB kinematics~\cite{Ahmad:2008hp,Goloskokov:2009ia,Goloskokov:2011rd,Goldstein:2012az}, or 
uncertainties in vector meson DAs in case of $\rho$- and $\phi$-meson production.

From this point of view, consistency checks of the GPD extraction procedure from
experimental data, especially of their flavor structure, are important.
Earlier we proposed to study the GPDs in deeply virtual neutrinoproduction
of the pseudo-Goldstone mesons ($\pi,\, K,\,\eta$)~\cite{Kopeliovich:2012dr}
with the high-intensity~\textsc{NuMI} beam at Fermilab, which recently
switched to the so-called middle-energy (ME) regime with an average
neutrino energy of about 6~GeV, and potentially is able to reach energies
up to 20 GeV, without essential loss of luminosity. The $\nu$DVMP
measurements with neutrino and antineutrino beams are complementary
to the electromagnetic DVMP. In the axial channel, due to the chiral
symmetry breaking we have an octet of pseudo-Goldstone bosons, which
act as a natural probe of the flavor content. Due to the $V-A$ structure
of the charged current, in $\nu$DVMP one can access simultaneously
the unpolarized GPDs, $H,\, E$, and the helicity flip GPDs, $\tilde{H}$
and $\tilde{E}$. Besides, using chiral symmetry and assuming closeness
of parameters of pion and kaon, full flavour structure of the GPDs
can be extracted.

It is worth reminding that the cross-sections were evaluated  in~\cite{Kopeliovich:2012dr}
in the leading twist approximation, and for a correct extraction of the
GPDs at the energies of MINERvA in ME regime, an estimate of the higher
twist effects is required. The first twist-3 correction arises due to
contribution of the transversely polarized intermediate virtual bosons and is
controlled by convolution of poorly known transversity GPDs $H_{T},\, E_{T},\,\tilde{H}_{T},\,\tilde{E}_{T}$
and twist-3 DAs of pion. While this correction vanishes
at asymptotically large $Q^{2}$, at moderate $Q^{2}$ in \emph{electro}production
it gives a sizable contribution, which was confirmed
in the CLAS experiment~\cite{Kubarovsky:2011zz}. In case of neutrino-production
situation is different since due to $V-A$ structure of the weak currents
there is an additional and numerically dominant contribution of the
unpolarized GPDs $H,\, E$ to the leading twist amplitude. In this
paper we analyze the relative size of the twist-3 contributions to
the neutrino-production of pions and demonstrate that they are indeed
small. In this respect we differ from~\cite{Goldstein:2009in}, where
the contribution of chiral odd GPDs was assumed to be numerically
dominant.

The paper is organized as follows. In Section~\ref{sec:DVMP_Xsec}
we evaluate the Goldstone meson production by neutrinos on nucleon
targets accounting for higher twist effects. In Section~\ref{sec:Parametrizations},
for the sake of completeness we highlight the properties of the GPD parametrization
used for evaluations. In Section~\ref{sec:Results}
we present numerical results and make conclusions.

\section{Cross-section of the $\nu$DVMP process}

\label{sec:DVMP_Xsec}The cross-section of the Goldstone mesons production
in  neutrino-hadron collisions has the form
\begin{align}
\frac{d\sigma}{dt\, dx_{B}dQ^{2}d\phi} & =\epsilon\frac{d\sigma_{L}}{dt\, dx_{B}dQ^{2}d\phi}+\frac{d\sigma_{T}}{dt\, dx_{B}dQ^{2}d\phi}+\sqrt{\epsilon(1+\epsilon)}\cos\phi\frac{d\sigma_{LT}}{dt\, dx_{B}dQ^{2}d\phi}\label{eq:sigma_def}\\
 & +\epsilon\cos2\phi\frac{d\sigma_{TT}}{dt\, dx_{B}dQ^{2}d\phi}+\sqrt{\epsilon(1+\epsilon)}\sin\phi\frac{d\sigma_{L'T}}{dt\, dx_{B}dQ^{2}d\phi}+\epsilon\sin2\phi\frac{d\sigma_{T'T}}{dt\, dx_{B}dQ^{2}d\phi},\nonumber 
\end{align}
where $t=\left(p_{2}-p_{1}\right)^{2}$ is the momentum transfer to
baryon, $Q^{2}=-q^{2}$ is the virtuality of the charged boson, $x_{B}=Q^{2}/(2p\cdot q)$
is Bjorken $x$, $\phi$ is the angle between the lepton and meson
production scattering planes, and we introduced shorthand notations
\[
\epsilon=\frac{1-y-\frac{\gamma^{2}y^{2}}{4}}{1-y+\frac{y^{2}}{2}+\frac{\gamma^{2}y^{2}}{4}},\quad\gamma=\frac{2\, m_{N}x_{B}}{Q},\quad y=\frac{Q^{2}}{sx_{B}}.
\]
 In the asymptotic Bjorken limit the cross-section is dominated by
the first angular independent term $\epsilon\, d\sigma_{L}/dt\, dx_{B}dQ^{2}d\phi$
which was studied in our previous paper~\cite{Kopeliovich:2012dr}
and is a straightforward extension of the electroproduction of pions
studied in~\cite{Vanderhaeghen:1998uc,Mankiewicz:1998kg,Goloskokov:2006hr,Goloskokov:2007nt,Goloskokov:2008ib,Goloskokov:2011rd,Goldstein:2012az}.
As we will see below the twist-3 corrections are small, for this reason
it is convenient to normalize all the cross-sections in~(\ref{eq:sigma_def})
to this term,
\begin{equation}
\frac{d\sigma}{dt\, dx_{B}dQ^{2}d\phi}=\epsilon\frac{d\sigma_{L}}{dt\, dx_{B}dQ^{2}d\phi}\sum_{n}\left(c_{n}\cos n\phi+s_{n}\sin n\phi\right)\label{eq:harmonicsDefinition}
\end{equation}
and discuss higher-twist effects in terms of harmonics $c_{n},\, s_{n}$.
In what follows, it is convenient to introduce a photon helicity matrix
$\sigma_{\alpha\beta}$ defined as

\begin{equation}
\sigma_{\alpha\beta}=\frac{1}{2}\sum_{\nu\nu'}\mathcal{A}_{\nu'0,\nu\alpha}^{*}\mathcal{A}_{\nu'0,\nu\beta},\label{eq:sigma_ij}
\end{equation}
where $\mathcal{A}_{\nu'0,\nu\alpha}$ is the amplitude of the corresponding
process in helicity basis, and $\nu,\,\nu'$ are the polarizations
of the initial and final baryon. In terms of $\sigma_{\alpha\beta}$
the cross-sections in~(\ref{eq:sigma_def}) can be written as

\begin{align}
\frac{d\sigma_{L}}{dt\, dx_{B}dQ^{2}d\phi} & =\Gamma\,\sigma_{00}\label{eq:sigma_L}\\
\frac{d\sigma_{T}}{dt\, dx_{B}dQ^{2}d\phi} & =\Gamma\,\left(\frac{\sigma_{++}+\sigma_{--}}{2}-\mu\sqrt{1-\epsilon^{2}}\frac{\sigma_{++}-\sigma_{--}}{2}\right)\label{eq:sigma_T}\\
\frac{d\sigma_{LT}}{dt\, dx_{B}dQ^{2}d\phi} & =\Gamma\,\left({\rm Re}\left(\sigma_{0+}-\sigma_{0-}\right)-\mu\sqrt{\frac{1-\epsilon}{1+\epsilon}}{\rm Re}\left(\sigma_{0+}+\sigma_{0-}\right)\right)\label{eq:sigma_LT}\\
\frac{d\sigma_{TT}}{dt\, dx_{B}dQ^{2}d\phi} & =-\Gamma\,{\rm Re}\left(\sigma_{+-}\right)\label{eq:sigma_TT}\\
\frac{d\sigma_{L'T}}{dt\, dx_{B}dQ^{2}d\phi} & =-\Gamma\,\left({\rm Im}\left(\sigma_{+0}+\sigma_{-0}\right)+\mu\sqrt{\frac{1-\epsilon}{1+\epsilon}}{\rm Im}\left(\sigma_{-0}-\sigma_{+0}\right)\right)\label{eq:sigma_LPrimeT}\\
\frac{d\sigma_{T'T}}{dt\, dx_{B}dQ^{2}d\phi} & =-\Gamma\,{\rm Im}\left(\sigma_{+-}\right)\label{eq:sigma_TPrimeT}
\end{align}

where we introduced shorthand notations $\Gamma$ and $\mu$, which
for the charged current (CC) and neutral current (NC) are defined
as 
\begin{align}
\Gamma_{CC} & =\frac{G_{F}^{2}f_{M}^{2}x_{B}^{2}\left(1-y+\frac{y^{2}}{2}+\frac{\gamma^{2}y^{2}}{4}\right)}{64\pi^{4}Q^{2}\left(1+Q^{2}/M_{W}^{2}\right)^{2}\left(1+\gamma^{2}\right)^{3/2}},\\
\Gamma_{NC} & =\frac{G_{F}^{2}f_{M}^{2}x_{B}^{2}\left(1-y+\frac{y^{2}}{2}+\frac{\gamma^{2}y^{2}}{4}\right)}{64\pi^{4}\cos^{4}\theta_{W}Q^{2}\left(1+Q^{2}/M_{Z}^{2}\right)^{2}\left(1+\gamma^{2}\right)^{3/2}},\\
\mu_{\nu,\bar{\nu}} & =\pm\frac{1}{2},\label{eq:mu_def}
\end{align}
$f_\pi$ is the pion decay constant, $G_F$ is the Fermi constant, $\theta_W$ is the Weinberg angle, and $M_W,\,M_Z$ are the masses of the $W$ and $Z$ bosons.

From~(\ref{eq:sigma_L}-\ref{eq:sigma_TPrimeT}) we can see that
the terms $\sim\sin\phi$ and $\sim\cos\phi$ appear due to interference
of the leading twist and twist-3 contributions, whereas all the other
contributions appear entirely due to the twist-3 effects. Since the twist-3
amplitudes are suppressed by $1/Q$ compared to the leading twist
result, we expect that in the large-$Q$ limit 
\begin{equation}
c_{1},\, s_{1}\sim1/Q,\quad c_{0}-1,c_{2},s_{2}\sim1/Q^{2}.
\end{equation}

Due to the factorization theorem the amplitude~$\mathcal{A}_{\nu'0,\nu\beta}$
in~(\ref{eq:sigma_ij}) may be written as a convolution of the hard and
soft parts,
\begin{equation}
\mathcal{A}_{\nu'0,\nu\alpha}=\int_{-1}^{+1}dx\sum_{q,q'=u,d,s}\sum_{\lambda\lambda'}\mathcal{H}_{\nu'\lambda',\nu\lambda}^{q'q}\mathcal{C}_{\lambda'0,\lambda\alpha}^{q'q},\label{eq:M_conv}
\end{equation}
where $x$ is the average light-cone fraction of the parton, $\lambda,\, q$
($\lambda',\, q'$) are the corresponding helicity and flavour of
the initial (final) partons, the helicity amplitude, $\mathcal{H}_{\nu'\lambda',\nu\lambda}^{q'q}$
is the process- and baryon-dependent soft part which will be specified
later, and $\mathcal{C}_{\lambda'\nu',\lambda\nu}^{q}$ is the hard
coefficient function.

In the leading twist, four GPDs, $H^{q'q},\, E^{q'q},\,\tilde{H}^{q'q}$
and $\tilde{E}^{q'q}$ contribute to $\mathcal{H}_{\nu'\lambda',\nu\lambda}^{q'q}$.
They are defined as 
\begin{eqnarray}
\frac{\bar{P}^{+}}{2\pi}\int dz\, e^{ix\bar{P}^{+}z}\left\langle B\left(p_{2}\right)\left|\bar{\psi}_{q'}\left(-\frac{z}{2}\right)\gamma_{+}\psi_{q}\left(\frac{z}{2}\right)\right|A\left(p_{1}\right)\right\rangle  & = & \left(H^{q'q}\left(x,\xi,t\right)\bar{N}\left(p_{2}\right)\gamma_{+}N\left(p_{1}\right)\right.\label{eq:H_def}\\
 &  & \left.+\frac{\Delta_{k}}{2m_{N}}E^{q'q}\left(x,\xi,t\right)\bar{N}\left(p_{2}\right)i\sigma_{+k}N\left(p_{1}\right)\right)\nonumber \\
\frac{\bar{P}^{+}}{2\pi}\int dz\, e^{ix\bar{P}^{+}z}\left\langle B\left(p_{2}\right)\left|\bar{\psi}_{q'}\left(-\frac{z}{2}\right)\gamma_{+}\gamma_{5}\psi_{q}\left(\frac{z}{2}\right)\right|A\left(p_{1}\right)\right\rangle  & = & \left(\tilde{H}^{q'q}\left(x,\xi,t\right)\bar{N}\left(p_{2}\right)\gamma_{+}\gamma_{5}N\left(p_{1}\right)\right.\label{eq:Htilde_def}\\
 &  & \left.+\frac{\Delta_{+}}{2m_{N}}\tilde{E}^{q'q}\left(x,\xi,t\right)\bar{N}\left(p_{2}\right)N\left(p_{1}\right)\right),\nonumber 
\end{eqnarray}
where $\bar{P}=p_{1}+p_{2}$, $\Delta=p_{2}-p_{1}$ and $\xi=-\Delta^{+}/2\bar{P}^{+}\approx x_{Bj}/(2-x_{Bj})$
(see e.g.~\cite{Goeke:2001tz} for the details of kinematics). In the
case when the baryon remains intact, $A=B$, the corresponding
GPDs are diagonal in the flavour space, $H^{q'q}\sim\delta_{q'q}H^{q}$,
etc. In the general case, when $A\not=B$, in the right-hand side
(r.h.s.) of Eqs.~(\ref{eq:H_def}), (\ref{eq:Htilde_def}) there
might be extra structures which otherwise are forbidden by $T$-parity in the
case of $A=B$~\cite{Goeke:2001tz}. In what follows we assume that
the target $A$ is either a proton or a neutron, and the recoil $B$
belongs to the same lowest $SU(3)$ octet of baryons. In this case,
all such terms are parametrically suppressed by the current quark
mass $m_{q}$ and vanish in the limit of exact $SU(3)$, so we will
disregard them. In this special case, we can rely on the $SU(3)$ relations
and express the nondiagonal transitional GPDs as linear combinations
of the GPDs of the proton $H^{q},\, E^{q},\,\tilde{H}^{q},\,\tilde{E}^{q}$~\cite{Frankfurt:1999fp},
so~(\ref{eq:M_conv}) may be effectively rewritten as 
\begin{equation}
\mathcal{A}_{\nu'0,\nu\alpha}=\int_{-1}^{+1}dx\sum_{q}\sum_{\lambda\lambda'}\mathcal{H}_{\nu'\lambda',\nu\lambda}^{q}\mathcal{C}_{\lambda'0,\lambda\alpha}^{q},\label{eq:M_conv_2}
\end{equation}

The twist-3 correction is controlled by the chiral odd transversity
GPDs defined as~\cite{Diehl:2001pm} 
\begin{eqnarray}
 &  & \frac{\bar{P}^{+}}{2\pi}\int dz\, e^{ix\bar{P}^{+}z}\left\langle B\left(p_{2}\right)\left|\bar{\psi}_{q'}\left(-\frac{z}{2}\right)i\sigma^{+j}\psi_{q}\left(\frac{z}{2}\right)\right|A\left(p_{1}\right)\right\rangle =\label{eq:HT_def}\\
 & = & \left(H_{T}^{q'q}\left(x,\xi,t\right)\bar{N}\left(p_{2}\right)i\sigma^{+j}N\left(p_{1}\right)+\tilde{H}_{T}^{q'q}\frac{\bar{P}^{+}\Delta^{j}-\Delta^{+}\bar{P}^{j}}{m^{2}}+E_{T}^{q'q}\frac{\gamma^{+}\Delta^{j}-\Delta^{+}\gamma^{j}}{2m}+\tilde{E}_{T}^{q'q}\frac{\gamma^{+}\bar{P}^{j}-\bar{P}^{+}\gamma^{j}}{m}\right),\nonumber 
\end{eqnarray}
where $j=1,2$ is the transverse index. Similar to the leading-twist
case, the flavour structure may be simplified with the help of $SU(3)$
relations and rewritten in terms of the transversity GPDs of the proton.

The matrix $\mathcal{H}_{\nu'\lambda',\nu\lambda}^{q}$ in~(\ref{eq:M_conv_2})
is a linear combination of the helicity-odd and even GPDs,
\begin{align}
\mathcal{H}_{\nu'\lambda',\nu\lambda}^{q} & =\frac{2\delta_{\lambda\lambda'}}{\sqrt{1-\xi^{2}}}\left(-\left(\begin{array}{cc}
\left(1-\xi^{2}\right)H^{q}-\xi^{2}E^{q} & \frac{\left(\Delta_{1}+i\Delta_{2}\right)E^{q}}{2m}\\
-\frac{\left(\Delta_{1}-i\Delta_{2}\right)E^{q}}{2m} & \left(1-\xi^{2}\right)H^{q}-\xi^{2}E^{q}
\end{array}\right)_{\nu'\nu}\right.\\
 & +\left.{\rm sgn}(\lambda)\left(\begin{array}{cc}
-\left(1-\xi^{2}\right)\tilde{H}^{q}+\xi^{2}\tilde{E}^{q} & \frac{\left(\Delta_{1}+i\Delta_{2}\right)\xi\tilde{E}^{q}}{2m}\\
\frac{\left(\Delta_{1}-i\Delta_{2}\right)\xi\tilde{E}^{q}}{2m} & \left(1-\xi^{2}\right)\tilde{H}^{q}-\xi^{2}\tilde{E}^{q}
\end{array}\right)_{\nu'\nu}\right)+\nonumber \\
 & +\left(m_{\nu'\nu}^{q}\delta_{\lambda,-}\delta_{\lambda',+}+n_{\nu'\nu}^{q}\delta_{\lambda,+}\delta_{\lambda',-}\right),\nonumber 
\end{align}
where the coefficients $m_{\pm,\pm}^{q}$ and $n_{\pm,\pm}^{q}$
are given by 
\begin{align}
m_{--}^{q} & =\frac{\sqrt{-t'}}{4m}\left[2\tilde{H}_{T}^{q}\,+(1+\xi)E_{T}^{q}-(1+\xi)\tilde{E}_{T}^{q}\right],\\
m_{-+}^{q} & =\sqrt{1-\xi^{2}}\frac{t'}{4m^{2}}\tilde{H}_{T}^{q},\\
m_{+-}^{q} & =\sqrt{1-\xi^{2}}\left[H_{T}^{q}-\frac{\xi^{2}}{1-\xi^{2}}E_{T}^{q}+\frac{\xi}{1-\xi^{2}}\tilde{E}_{T}^{q}-\frac{t'}{4m^{2}}\tilde{H}_{T}^{q}\right],\\
m_{++}^{q} & =\frac{\sqrt{-t'}}{4m}\left[2\tilde{H}_{T}^{q}+(1-\xi)E_{T}^{q}+(1-\xi)\tilde{E}_{T}^{q}\right],
\end{align}

\begin{align}
n_{--}^{q} & =-\frac{\sqrt{-t'}}{4m}\left(2\tilde{H}_{T}^{q}+(1-\xi)E_{T}^{q}+(1-\xi)\tilde{E}_{T}^{q}\right),\\
n_{-+}^{q} & =\sqrt{1-\xi^{2}}\left(H_{T}^{q}-\frac{\xi^{2}}{1-\xi^{2}}E_{T}^{q}+\frac{\xi}{1-\xi^{2}}\tilde{E}_{T}^{q}-\frac{t'}{4m^{2}}\tilde{H}_{T}^{q}\right),\\
n_{+-}^{q} & =\sqrt{1-\xi^{2}}\frac{t'}{4m^{2}}\tilde{H}_{T}^{q},\\
n_{++}^{q} & =-\frac{\sqrt{-t'}}{4m}\left(2\tilde{H}_{T}^{q}+(1+\xi)E_{T}^{q}-(1+\xi)\tilde{E}_{T}^{q}\right),
\end{align}
and we introduced a shorthand notation $t'=-\Delta_\perp^2/(1-\xi^2)$, where $\Delta_\perp=p_{2,\perp}-p_{1,\perp}$ is the transverse part of the momentum transfer.

\begin{figure}[htp]
\includegraphics[scale=0.4]{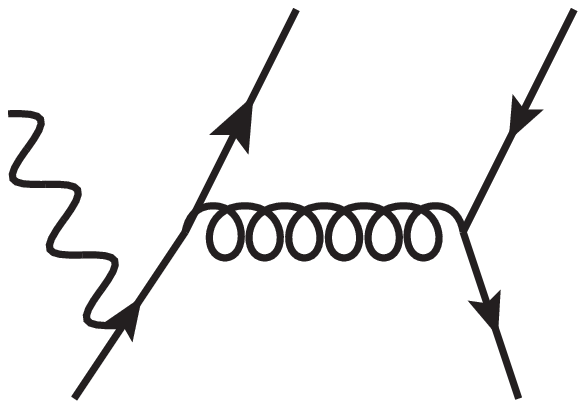}\includegraphics[scale=0.4]{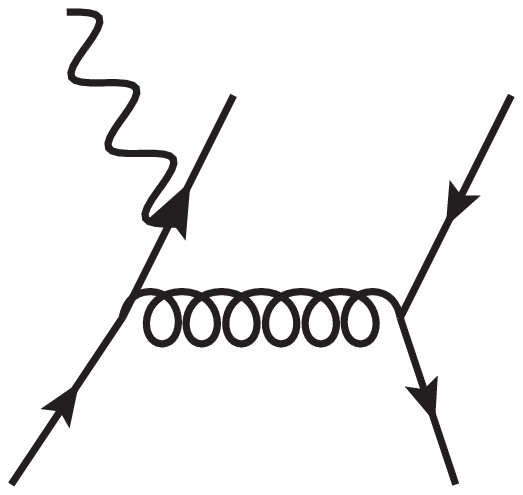}\includegraphics[scale=0.4]{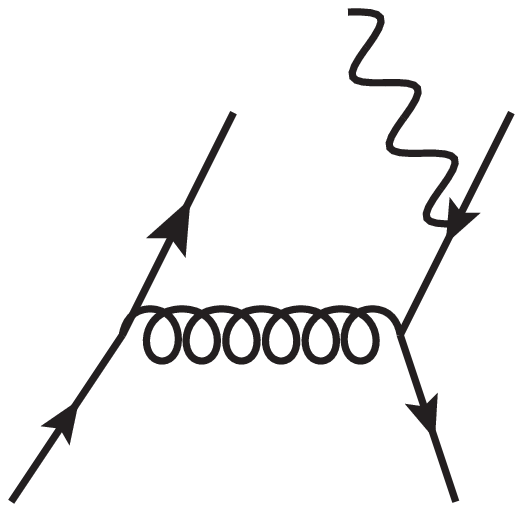}\includegraphics[scale=0.4]{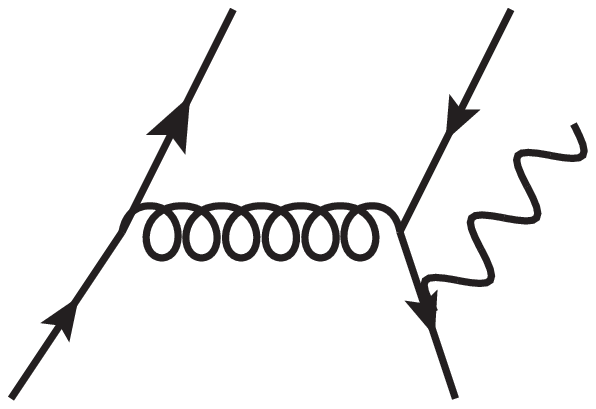}
\caption{\label{fig:DVMPLT-1}Leading-order contributions to the DVMP hard
coefficient functions.}
\end{figure}

Evaluation of the hard coefficient function $\mathcal{C}_{\lambda'0,\lambda\mu}^{q}$
is quite straightforward, and in the leading order over $\alpha_{s}$
is given by the four diagrams shown schematically in Figure~\ref{fig:DVMPLT-1}. 

A key ingredient in evaluation of the coefficient functions are the
distribution amplitudes (DAs) of the produced pion. Since we are interested
in making evaluations up to twist-3 effects, we have to take into account
both twist-2 and twist-3 pion DAs, defined respectively as~\cite{Kopeliovich:2011rv}
\begin{eqnarray}
\phi_{2}\left(z\right) & = & \frac{1}{if_{\pi}\sqrt{2}}\int\frac{du}{2\pi}e^{i(z-0.5)u}\left\langle 0\left|\bar{\psi}\left(-\frac{u}{2}n\right)\hat{n}\gamma_{5}\psi\left(\frac{u}{2}n\right)\right|\pi(q)\right\rangle ,\label{eq:DA2p}
\end{eqnarray}

\begin{eqnarray}
\phi_{3}^{(p)}\left(z\right) & = & \frac{1}{f_{\pi}\sqrt{2}}\frac{m_{u}+m_{d}}{m_{\pi}^{2}}\int\frac{du}{2\pi}e^{i(z-0.5)u}\left\langle 0\left|\bar{\psi}\left(-\frac{u}{2}n\right)\gamma_{5}\psi\left(\frac{u}{2}n\right)\right|\pi(q)\right\rangle ,\label{eq:DA3p}
\end{eqnarray}

\begin{eqnarray}
\phi_{3}^{(\sigma)}\left(z\right) & = & \frac{3i}{\sqrt{2}f_{\pi}}\frac{m_{u}+m_{d}}{m_{\pi}^{2}}\int\frac{du}{2\pi}e^{i(z-0.5)u}\left\langle 0\left|\bar{\psi}\left(-\frac{u}{2}n\right)\sigma_{+-}\gamma_{5}\psi\left(\frac{u}{2}n\right)\right|\pi(q)\right\rangle .\label{eq:DA3s}
\end{eqnarray}
As one can see from~(\ref{eq:DA2p}-\ref{eq:DA3s}), the pseudovector
and pseudoscalar DAs $\phi_{2;\pi},\,\phi_{3;\pi}^{(p)}$ are chiral
even (symmetric w.r.t. $z\to1-z$), whereas the tensor DA $\phi_{3;\pi}^{(\sigma)}$
is chiral odd. Straightforward evaluation of the diagrams shown in
Figure~\ref{fig:DVMPLT-1} yields for the coefficient function
\begin{align}
\mathcal{C}_{\lambda'0,\lambda\mu}^{q} & =\delta_{\mu0}\delta_{\lambda\lambda'}\sum_{k=\pm}\left[\eta_{A,k}^{q}c_{k}^{(2)}\left(x,\xi\right)+{\rm sgn}(\lambda)\eta_{V,k}^{q}c_{k}^{(2)}\left(x,\xi\right)\right]+\\
 & +\delta_{\mu,+}\delta_{\lambda,-}\delta_{\lambda',+}\left(S_{A}^{q}-S_{V}^{q}\right)+\delta_{\mu,-}\delta_{\lambda,+}\delta_{\lambda',-}\left(S_{A}^{q}+S_{V}^{q}\right)+\mathcal{O}\left(\frac{m^{2}}{Q^{2}}\right), \label{eq:Coef_function}
\end{align}
where we introduced shorthand notations 
\begin{eqnarray}
S_{A}^{q} & = & \int dz\,\left(\left(\eta_{A+}^{q}c_{+}^{(3,p)}\left(x,\xi\right)-\eta_{A-}^{q}c_{-}^{(3,p)}\left(x,\xi\right)\right)+2\left(\eta_{A-}^{q}c_{-}^{(3,\sigma)}\left(x,\xi\right)+\eta_{A+}^{q}c_{+}^{(3,\sigma)}\left(x,\xi\right)\right)\right),\label{eq:SA_def}\\
S_{V}^{q} & = & \int dz\,\left(\left(\eta_{V+}^{q}c_{+}^{(3,p)}\left(x,\xi\right)+\eta_{V-}^{q}c_{-}^{(3,p)}\left(x,\xi\right)\right)+2\left(\eta_{V+}^{q}c_{+}^{(3,\sigma)}\left(x,\xi\right)-\eta_{V-}^{q}c_{-}^{(3,\sigma)}\left(x,\xi\right)\right)\right),\label{eq:SV_def}\\
c_{\pm}^{(2)}\left(x,\xi\right) & = & \left(\int dz\frac{\phi_{2}(z)}{z}\right)\frac{8\pi i}{9}\frac{\alpha_{s}f_{\pi}}{Q}\frac{1}{x\pm\xi\mp i0},
\end{eqnarray}

\begin{equation}
c_{+}^{(3,i)}\left(x,\xi\right)=\frac{4\pi i\alpha_{s}f_{\pi}\xi}{9\, Q^{2}}\int_{0}^{1}dz\frac{\phi_{3,i}(z)}{z(x+\xi)^{2}},\quad c_{-}^{(3,i)}\left(x,\xi\right)=\frac{4\pi i\alpha_{s}f_{\pi}\xi}{9\, Q^{2}}\int_{0}^{1}dz\frac{\phi_{3,i}(z)}{(1-z)(x-\xi)^{2}};\label{eq:Tw3_coefFunction}
\end{equation}
and the process-dependent flavor factors $\eta_{V\pm}^{q},\,\eta_{A\pm}^{q}$
are presented in  table~\ref{tab:DVMP_amps}. As was discussed above,
for the processes, in which either initial or final baryon is different
from proton, we used $SU(3)$ relations~\cite{Frankfurt:1999fp}, 
valid up to the corrections in current quark mass $\sim\mathcal{O}\left(m_{q}\right)$.
In the leading twist, due to the underlying $SU(3)$ relations there are
identities relating the neutrino-hadron and antineutrino-hadron cross-sections.
In the next-to-leading order these relations are broken due to the
weak isospin-dependent factor $\mu$ in~(\ref{eq:mu_def}), and corresponding
$SU(3)$ identities are valid only for the cross-sections $d\sigma_{TT}$
and $d\sigma_{T'T}$ (harmonics $c_{2},\, s_{2}$). For certain processes
either $\eta_{+}^{q}$ or $\eta_{-}^{q}$ might vanish. In this case
from the definition~(\ref{eq:SA_def},\ref{eq:SV_def}) we may see
that $S_{A}=\pm S_{V}$, and the matrix element $\sigma_{+-}$ which
controls the cross-sections~$d\sigma_{TT},\, d\sigma_{T'T}$~(\ref{eq:sigma_TT},\ref{eq:sigma_TPrimeT})
vanishes identically. 

\begin{table}[h]
\caption{\label{tab:DVMP_amps}The flavour coefficients $\eta_{\pm}^{q}$ for
various processes ($q=u,d,s,...$). For the case of CC mediated processes,
take $\eta_{V\pm}^{q}=\eta_{\pm}^{q},\quad\eta_{A\pm}^{q}=-\eta_{\pm}^{q}$.
For the case of NC mediated processes, take $g_{q}$ corresponding
to vector current coupling $g_{V}^{q}$ and axial-vector current coupling $g_{A}^{q}$ for the helicity odd and even
GPDs respectively.}

\global\long\def\arraystretch{1.5}

\begin{tabular}{|c|c|c|c|c|c|c|c|c|}
\cline{1-4} \cline{6-9} 
Process  & type  & $\eta_{+}^{q}$ & $\eta_{-}^{q}$ &  & Process  & type  & $\eta_{+}^{q}$ & $\eta_{-}^{q}$\tabularnewline
\cline{1-4} \cline{6-9} 
$\nu\, p\to\mu^{-}\pi^{+}p$  & CC  & $V_{ud}\delta_{qu}$ & $V_{ud}\delta_{qd}$ &  & $\nu\, n\to\mu^{-}\pi^{+}n$  & CC  & $V_{ud}\delta_{qd}$ & $V_{ud}\delta_{qu}$\tabularnewline
\cline{1-4} \cline{6-9} 
$\bar{\nu}\, p\to\mu^{+}\pi^{-}p$  & CC  & $V_{ud}\delta_{qd}$ & $V_{ud}\delta_{qu}$ &  & $\bar{\nu}\, n\to\mu^{+}\pi^{-}n$  & CC  & $V_{ud}\delta_{qu}$ & $V_{ud}\delta_{qd}$\tabularnewline
\cline{1-4} \cline{6-9} 
$\bar{\nu\,}p\to\mu^{+}\pi^{0}n$  & CC  & $V_{ud}\frac{\delta_{qu}-\delta_{qd}}{\sqrt{2}}$ & $-V_{ud}\frac{\delta_{qu}-\delta_{qd}}{\sqrt{2}}$ &  & $\nu\, n\to\mu^{-}\pi^{0}p$  & CC  & $-V_{ud}\frac{\delta_{qu}-\delta_{qd}}{\sqrt{2}}$ & $V_{ud}\frac{\delta_{qu}-\delta_{qd}}{\sqrt{2}}$\tabularnewline
\cline{1-4} \cline{6-9} 
$\nu\, p\to\nu\,\pi^{+}n$  & NC  & $g_{d}\left(\delta_{qu}-\delta_{qd}\right)$ & $g_{u}\left(\delta_{qu}-\delta_{qd}\right)$ &  & $\nu\, n\to\nu\,\pi^{-}p$  & NC  & $g_{u}\left(\delta_{qu}-\delta_{qd}\right)$ & $g_{d}\left(\delta_{qu}-\delta_{qd}\right)$\tabularnewline
\cline{1-4} \cline{6-9} 
$\nu\, p\to\nu\,\pi^{0}p$  & NC  & $\frac{g_{u}\delta_{qu}-g_{d}\delta_{qd}}{\sqrt{2}}$ & $\frac{g_{u}\delta_{qu}-g_{d}\delta_{qd}}{\sqrt{2}}$ &  & $\nu\, n\to\nu\,\pi^{0}n$  & NC  & $\frac{g_{u}\delta_{qd}-g_{d}\delta_{qu}}{\sqrt{2}}$ & $\frac{g_{u}\delta_{qd}-g_{d}\delta_{qu}}{\sqrt{2}}$\tabularnewline
\cline{1-4} \cline{6-9} 
$\bar{\nu\,}p\to\mu^{+}\pi^{-}\Sigma_{+}$  & CC  & $-V_{us}\left(\delta_{qd}-\delta_{qs}\right)$ & 0 &  & $\bar{\nu\,}n\to\mu^{+}\pi^{-}\Lambda$  & CC  & $-V_{us}\frac{2\delta_{qd}-\delta_{qu}-\delta_{qs}}{\sqrt{6}}$ & 0\tabularnewline
\cline{1-4} \cline{6-9} 
$\bar{\nu\,}p\to\mu^{+}\pi^{0}\Sigma_{0}$  & CC  & $\frac{V_{us}}{2}\left(\delta_{qd}-\delta_{qs}\right)$ & 0 &  & $\bar{\nu\,}n\to\mu^{+}\pi^{-}\Sigma_{0}$  & CC  & $-V_{us}\frac{\delta_{qu}-\delta_{qs}}{\sqrt{2}}$ & 0\tabularnewline
\cline{1-4} \cline{6-9} 
$\bar{\nu\,}p\to\mu^{+}\pi^{0}\Lambda$  & CC  & $V_{us}\frac{2\delta_{qu}-\delta_{qd}-\delta_{qs}}{2\sqrt{3}}$ & 0 &  & $\bar{\nu\,}n\to\mu^{+}\pi^{0}\Sigma^{-}$  & CC  & $V_{us}\frac{\delta_{qu}-\delta_{qs}}{\sqrt{2}}$ & 0\tabularnewline
\cline{1-4} \cline{6-9} 
\multicolumn{1}{c}{} & \multicolumn{3}{c}{} & \multicolumn{1}{c}{} & \multicolumn{1}{c}{} & \multicolumn{3}{c}{}\tabularnewline
\cline{1-4} \cline{6-9} 
$\nu\, p\to\mu^{-}K^{+}p$  & CC  & $V_{us}\delta_{qu}$ & $V_{us}\delta_{qs}$ &  & $\nu\, n\to\mu^{-}K^{+}n$  & CC  & $V_{us}\delta_{qd}$ & $V_{us}\delta_{qs}$\tabularnewline
\cline{1-4} \cline{6-9} 
$\bar{\nu}\, p\to\mu^{+}K^{-}p$  & CC  & $V_{us}\delta_{qs}$ & $V_{us}\delta_{qu}$ &  & $\bar{\nu}\, n\to\mu^{+}K^{-}n$  & CC  & $V_{us}\delta_{qs}$ & $V_{us}\delta_{qd}$\tabularnewline
\cline{1-4} \cline{6-9} 
$\bar{\nu}\, p\to\mu^{+}K^{0}\Sigma_{0}$  & CC  & 0 & $-V_{ud}\frac{\delta_{qd}-\delta_{qs}}{\sqrt{2}}$ &  & $\bar{\nu}\, n\to\mu^{+}K^{0}\Sigma^{-}$  & CC  & 0 & $-V_{ud}\left(\delta_{qu}-\delta_{qs}\right)$\tabularnewline
\cline{1-4} \cline{6-9} 
$\bar{\nu}\, p\to\mu^{+}K^{0}\Lambda$  & CC  & 0 & $-V_{ud}\frac{2\delta_{qu}-\delta_{qd}-\delta_{qs}}{\sqrt{6}}$ &  & $\nu\, n\to\nu K^{0}\Lambda$  & NC  & $-g_{d}\frac{2\delta_{qd}-\delta_{qu}-\delta_{qs}}{\sqrt{6}}$ & $-g_{d}\frac{2\delta_{qd}-\delta_{qu}-\delta_{qs}}{\sqrt{6}}$\tabularnewline
\cline{1-4} \cline{6-9} 
$\bar{\nu}\, p\to\mu^{+}\bar{K}^{0}n$  & CC  & 0 & $-V_{us}\left(\delta_{qu}-\delta_{qd}\right)$ &  & $\nu\, n\to\nu K^{0}\Sigma_{0}$  & NC  & $-g_{d}\frac{\delta_{qu}-\delta_{qs}}{\sqrt{2}}$ & $-g_{d}\frac{\delta_{qu}-\delta_{qs}}{\sqrt{2}}$\tabularnewline
\cline{1-4} \cline{6-9} 
$\nu\, p\to\mu^{-}K^{+}\Sigma^{+}$  & CC  & 0 & $-V_{ud}\left(\delta_{qd}-\delta_{qs}\right)$ &  & $\nu\, n\to\mu^{-}K^{+}\Sigma^{0}$  & CC  & 0 & $-V_{ud}\frac{\delta_{qu}-\delta_{qs}}{\sqrt{2}}$\tabularnewline
\cline{1-4} \cline{6-9} 
$\nu\, p\to\nu\, K^{+}\Lambda$  & NC  & $-g_{d}\frac{2\delta_{qu}-\delta_{qd}-\delta_{qs}}{\sqrt{6}}$ & $-g_{u}\frac{2\delta_{qu}-\delta_{qd}-\delta_{qs}}{\sqrt{6}}$ &  & $\nu\, n\to\mu^{-}\, K^{+}\Lambda$  & CC  & 0 & $-V_{ud}\frac{2\delta_{qd}-\delta_{qu}-\delta_{qs}}{\sqrt{6}}$\tabularnewline
\cline{1-4} \cline{6-9} 
$\nu\, p\to\nu\, K^{+}\Sigma_{0}$  & NC  & $g_{d}\frac{\delta_{qd}-\delta_{qs}}{\sqrt{2}}$ & $g_{u}\frac{\delta_{qd}-\delta_{qs}}{\sqrt{2}}$ &  & $\nu\, n\to\mu^{-}K^{0}p$  & CC  & $V_{us}\left(\delta_{qu}-\delta_{qd}\right)$ & $0$\tabularnewline
\cline{1-4} \cline{6-9} 
$\nu\, p\to\nu\, K^{0}\Sigma^{+}$  & NC  & $-g_{d}\left(\delta_{qd}-\delta_{qs}\right)$ & $-g_{d}\left(\delta_{qd}-\delta_{qs}\right)$ &  & $\nu\, n\to\nu\, K^{+}\Sigma^{-}$  & NC  & $-g_{d}\left(\delta_{qu}-\delta_{qs}\right)$ & $-g_{u}\left(\delta_{qu}-\delta_{qs}\right)$\tabularnewline
\cline{1-4} \cline{6-9} 
\multicolumn{1}{c}{} & \multicolumn{3}{c}{} & \multicolumn{1}{c}{} & \multicolumn{1}{c}{} & \multicolumn{3}{c}{}\tabularnewline
\cline{1-4} \cline{6-9} 
$\nu\, p\to\nu\,\eta\, p$  & NC  & $\frac{g_{u}\delta_{qu}+g_{d}\delta_{qd}-2g_{d}\delta_{qs}}{\sqrt{6}}$ & $\frac{g_{u}\delta_{qu}+g_{d}\delta_{qd}-2g_{d}\delta_{qs}}{\sqrt{6}}$ &  & $\nu\, n\to\nu\,\eta\, n$  & NC  & $\frac{g_{u}\delta_{qd}+g_{d}\delta_{qu}-2g_{d}\delta_{qs}}{\sqrt{6}}$ & $\frac{g_{u}\delta_{qd}+g_{d}\delta_{qu}-2g_{d}\delta_{qs}}{\sqrt{6}}$\tabularnewline
\cline{1-4} \cline{6-9} 
$\bar{\nu}\, p\to\mu^{+}\,\eta\, n$  & CC  & $V_{ud}\frac{\delta_{qu}-\delta_{qd}}{\sqrt{6}}$ & $V_{ud}\frac{\delta_{qu}-\delta_{qd}}{\sqrt{6}}$ &  & $\bar{\nu}\, n\to\mu^{+}\,\eta\,\Sigma^{-}$  & CC  & $-V_{us}\frac{\delta_{qu}-\delta_{qs}}{\sqrt{6}}$ & $2V_{us}\frac{\delta_{qu}-\delta_{qs}}{\sqrt{6}}$\tabularnewline
\cline{1-4} \cline{6-9} 
$\bar{\nu}\, p\to\mu^{+}\,\eta\,\Sigma_{0}$  & CC  & $V_{us}\frac{\delta_{qu}-\delta_{qd}}{2\sqrt{3}}$ & $-V_{us}\frac{\delta_{qu}-\delta_{qd}}{\sqrt{3}}$ &  & $\nu\, n\to\mu^{-}\,\eta\, p$  & CC  & $V_{ud}\frac{\delta_{qu}-\delta_{qd}}{\sqrt{6}}$ & $V_{ud}\frac{\delta_{qu}-\delta_{qd}}{\sqrt{6}}$\tabularnewline
\cline{1-4} \cline{6-9} 
$\bar{\nu}\, p\to\mu^{+}\,\eta\,\Lambda$  & CC  & $V_{us}\frac{2\delta_{qu}-\delta_{qd}-\delta_{qs}}{6}$ & $-V_{us}\frac{2\delta_{qu}-\delta_{qd}-\delta_{qs}}{3}$ &  &  &  &  & \tabularnewline
\cline{1-4} \cline{6-9} 
\end{tabular}
\end{table}

Using symmetry of $\phi_{p}$ and antisymmetry of $\phi_{\sigma}$
with respect to charge conjugation, we can show that dependence on
the pion DAs factorizes in the collinear approximation and contributes
only as the minus first moment of the linear combination of the twist-3
DAs, $\phi_{p}(z)+2\phi_{\sigma}(z)$, 
\begin{equation}
\left\langle \phi_{3}^{-1}\right\rangle =\int_{0}^{1}dz\frac{\phi_{3}^{(p)}\left(z\right)+2\phi_{3}^{(\sigma)}\left(z\right)}{z}.
\end{equation}
We see from~(\ref{eq:M_conv_2}, \ref{eq:Tw3_coefFunction}) that, excluding the very
special case when all the transversity GPDs vanish at $x=\pm\xi$,
the transverse amplitude suffers from a collinear singularity at these
two points. In order to regularize it, we follow~\cite{Goloskokov:2009ia}
and introduce a small transverse momentum of the quarks inside the meson.
Such regularization modifies~(\ref{eq:Tw3_coefFunction}) to
\begin{align}
c_{+}^{(3,i)}\left(x,\xi\right) & =\frac{4\pi i\alpha_{s}f_{\pi}\xi}{9\, Q^{2}}\int_{0}^{1}dz\, d^{2}l_{\perp}\frac{\phi_{3,i}\left(z,\, l_{\perp}\right)}{(x+\xi-i0)\left(z(x+\xi)+\frac{2\xi\, l_{\perp}^{2}}{Q^{2}}\right)},\label{eq:c3Plus}\\
c_{-}^{(3,i)}\left(x,\xi\right) & =\frac{4\pi i\alpha_{s}f_{\pi}\xi}{9\, Q^{2}}\int_{0}^{1}dz\, d^{2}l_{\perp}\frac{\phi_{3,i}\left(z,\, l_{\perp}\right)}{(x-\xi+i0)\left((1-z)(x-\xi)-\frac{2\xi\, l_{\perp}^{2}}{Q^{2}}\right)},\label{eq:c3Minus}
\end{align}
where $l_{\perp}$ is the transverse momentum of the quark, 
and we tacitly assume absence of any other transverse momenta
in the coefficient function. 

\section{GPD and DA parametrizations}

\label{sec:Parametrizations}The pion DAs are one of the main sources of uncertainty in the present analysis.
For the leading twist DA $\phi_{2\pi}(x)$, the currently available
data on meson photoproduction formfactor $F_{\pi\gamma\gamma}\left(Q^{2}\right)$
are compatible with an asymptotic form $\phi_{as}(z)=6\sqrt{2}f_{\pi}z(1-z)$,
with a typical uncertainty in the minus-first moment of the order
of $\sim10\%$ (see e.g.~\cite{Pimikov:2012nm,Bakulev:2012nh} and
reviews in~\cite{Brodsky:2011xx,Brodsky:2011yv}). 

For the twist-3 contribution, as was discussed in Section~\ref{sec:DVMP_Xsec},
the DAs $\phi_{3;p}\left(z,l_{\perp}\right)$ and $\phi_{3;\sigma}\left(z,l_{\perp}\right)$
contribute in a linear combination 
\begin{equation}
\phi_{3}\left(z,\, l_{\perp}\right)=\phi_{3;p}\left(z,\, l_{\perp}\right)+2\phi_{3;\sigma}\left(z,\, l_{\perp}\right).\label{eq:phi_3}
\end{equation}
For the sake of simplicity, we parametrize~(\ref{eq:phi_3}) in the
form
\begin{equation}
\phi_{3}\left(z,\, l_{\perp}\right)=\frac{2a_{p}^{3}}{\pi^{3/2}}l_{\perp}\phi_{as}(z)\exp\left(-a_{p}^{2}l_{\perp}^{2}\right),\label{eq:phi_3_Explicit}
\end{equation}
where the numerical constant $a_{p}$ is taken as $a_{p}\approx2\,{\rm GeV}^{-1}$
in analogy with~\cite{Goloskokov:2009ia,Goloskokov:2011rd}.

More than a dozen of different parametrizations of GPDs have been
proposed in the literature~\cite{Diehl:2000xz,Goloskokov:2008ib,Radyushkin:1997ki,Kumericki:2011rz,Kumericki:2009uq,Guidal:2010de,Polyakov:2008aa,Polyakov:2002wz,Freund:2002qf,Goldstein:2013gra,Goldstein:2014aja,Manohar:2010zm}.
While we neither endorse nor refute any of them, for the sake of concreteness
we use the parametrization~\cite{Goloskokov:2006hr,Goloskokov:2007nt,Goloskokov:2008ib},
which successfully described HERA~\cite{Aaron:2009xp} and JLAB~\cite{Goloskokov:2006hr,Goloskokov:2007nt,Goloskokov:2008ib}
data on electroproduction of different mesons, so is expected to provide
a reasonable description of $\nu$DVMP. The parametrization is based
on the Radyushkin's double distribution ansatz. It assumes additivity
of the valence and sea parts of the GPDs, 
\[
H(x,\xi,t)=H_{val}(x,\xi,t)+H_{sea}(x,\xi,t),
\]
which are defined as 
\begin{eqnarray}
H_{val}^{q} & = & \int_{|\alpha|+|\beta|\le1}d\beta d\alpha\delta\left(\beta-x+\alpha\xi\right)\,\frac{3\theta(\beta)\left((1-|\beta|)^{2}-\alpha^{2}\right)}{4(1-|\beta|)^{3}}q_{val}(\beta)e^{\left(b_{i}-\alpha_{i}\ln|\beta|\right)t},\label{eq:H_val}\\
H_{sea}^{q} & = & \int_{|\alpha|+|\beta|\le1}d\beta d\alpha\delta\left(\beta-x+\alpha\xi\right)\,\frac{3\, sgn(\beta)\left((1-|\beta|)^{2}-\alpha^{2}\right)^{2}}{8(1-|\beta|)^{5}}q_{sea}(\beta)e^{\left(b_{i}-\alpha_{i}\ln|\beta|\right)t},\label{eq:H_sea}
\end{eqnarray}
and $q_{val}$ and $q_{sea}$ are the ordinary valence and sea components
of PDFs. The coefficients $b_{i}$, $\alpha_{i}$, as well as the
parametrization of the input PDFs $q(x),\,\Delta q(x)$ and pseudo-PDFs
$e(x),\,\tilde{e}(x)$ (which correspond to the forward limit of the
GPDs $E,\,\tilde{E}$) are discussed in~\cite{Goloskokov:2006hr,Goloskokov:2007nt,Goloskokov:2008ib}.
The unpolarized PDFs $q(x)$ are adjusted to reproduce the CTEQ PDFs
in the limited range $4\lesssim Q^{2}\lesssim40$~GeV$^{2}$. Notice
that in this model the sea is flavor symmetric for asymptotically
large $Q^{2}$, 
\begin{equation}
H_{sea}^{u}=H_{sea}^{d}=\kappa\left(Q^{2}\right)H_{sea}^{s},\label{eq:SeaFlavourSymmetry}
\end{equation}
where 
\[
\kappa\left(Q^{2}\right)=1+\frac{0.68}{1+0.52\ln\left(Q^{2}/Q_{0}^{2}\right)},\quad Q_{0}^{2}=4\, GeV^{2}.
\]

The equality of the sea components of the light quarks in~(\ref{eq:SeaFlavourSymmetry})
should be considered only as a rough approximation, since in the forward
limit the inequality $\bar{d}\not=\bar{u}$ was firmly established
by the E866/NuSea experiment~\cite{Hawker:1998ty}. For this reason,
the predictions made with this parametrization of the GPDs for the $p\rightleftarrows n$
transitions in the region $x_{Bj}\in(0.1...0.3)$ might slightly underestimate
the data. 

The transversity GPDs in the parametrization~\cite{Goloskokov:2011rd}
are obtained using a familiar double-distribution based parametrization~(\ref{eq:H_val},\ref{eq:H_sea}),
and the forward limit of these GPDs is parametrized as 
\begin{align*}
H_{T}^{a}(x,0,0) & =N_{H_{T}}^{a}\sqrt{x}(1-x)\left(q^{a}(x)+\Delta q^{a}(x)\right),\\
\bar{E}_{T}(x,0,0) & \equiv E_{T}(x,0,0)+2\tilde{H}_{T}(x,0,0)=N_{\bar{E}_{T}}^{a}x^{-\alpha}(1-x)^{\beta},
\end{align*}
where the values of the parameters $N_{i}^{a},\alpha,\beta$ are fixed
from the lattice data. Since in the parametrization~\cite{Goloskokov:2011rd},
as well as in any other parametrization of chiral-odd GPDs available in
the literature, $s$-quarks are not included, we do not make any predictions
for strangeness production.

\section{Numerical results and discussion}

\label{sec:Results}

In this section we would like to present numerical results for the
twist-3 corrections to pion production using the Kroll-Goloskokov
parametrization of GPDs~\cite{Goloskokov:2006hr,Goloskokov:2007nt,Goloskokov:2008ib,Goloskokov:2009ia},
briefly discussed in the section~\ref{sec:Parametrizations}. As
was discussed in section~\ref{sec:DVMP_Xsec}, one of the consequences
of the higher twist corrections is the appearance of the azimuthal
angular dependence of the DVMP cross-sections~%
Since the twist-3 corrections are small, in what follows we prefer
to discuss the results in terms of the angular harmonics $c_{n},\, s_{n}$
defined in~(\ref{eq:harmonicsDefinition}). The most important harmonics
is $c_{0}$, because its deviation from unity affects the extraction
of GPDs in the leading twist approximation, and extraction of the leading twist result requires the experimentally challenging Rosenbluth separation with varying energy neutrino beam.
All the other harmonics generate nontrivial angular dependence and can be easily separated from the leading-twist contribution. For example, the
angle-integrated cross-section $d\sigma/d\ln x_{B}dt\, dQ^{2}$ is not sensitive to those harmonics at all.

\begin{figure}
\includegraphics[scale=0.3]{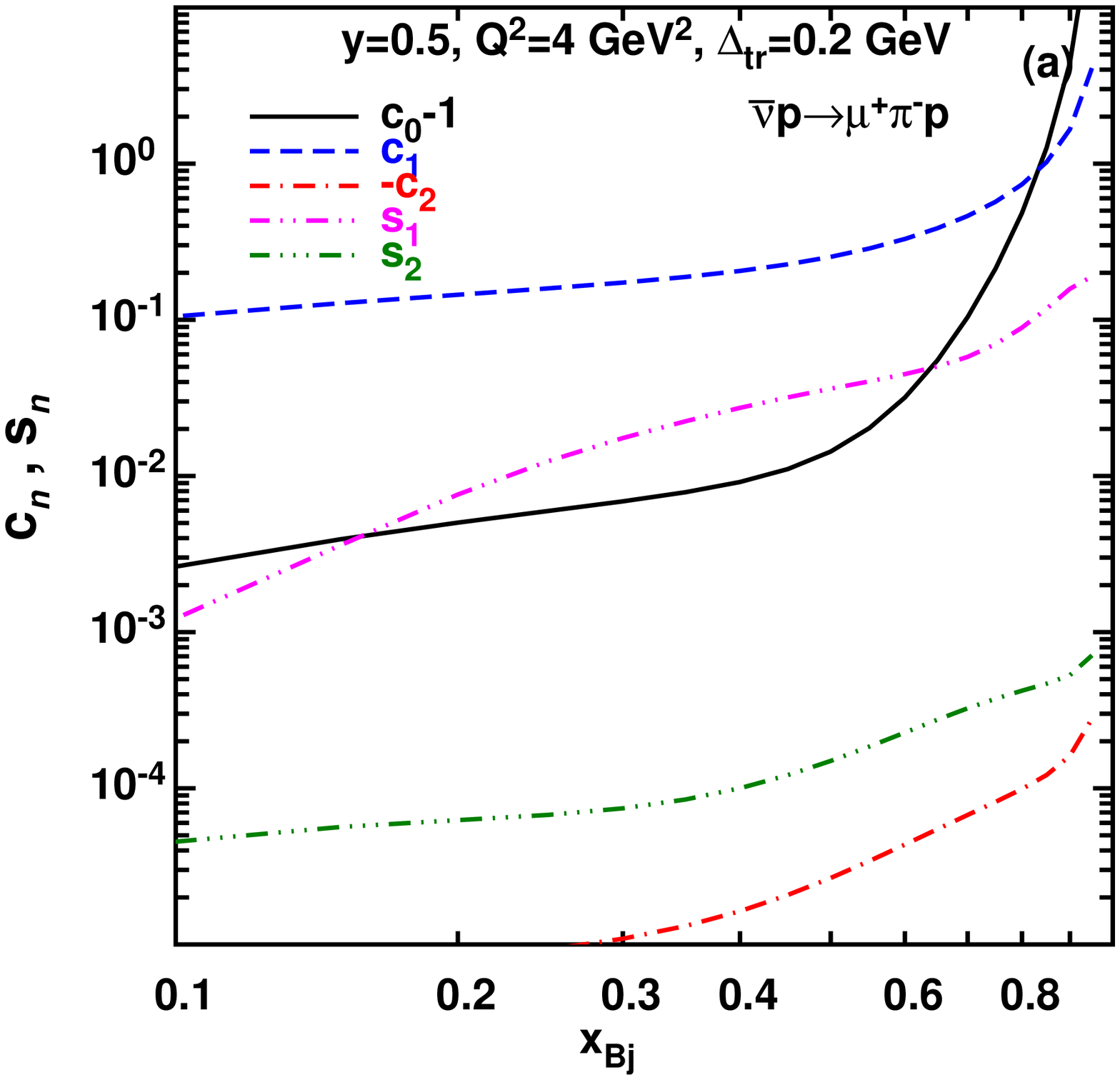}\includegraphics[scale=0.3]{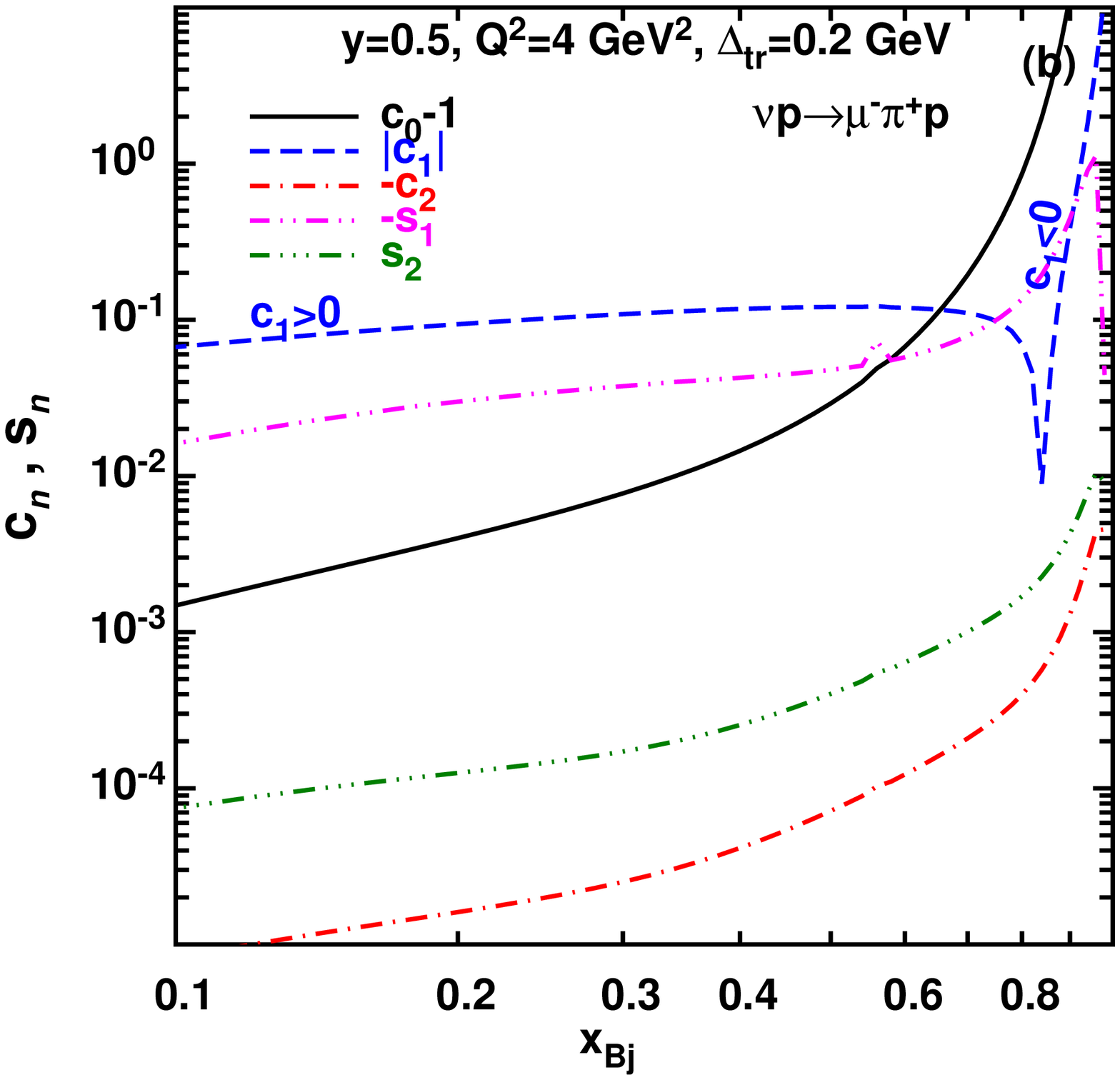}\includegraphics[scale=0.3]{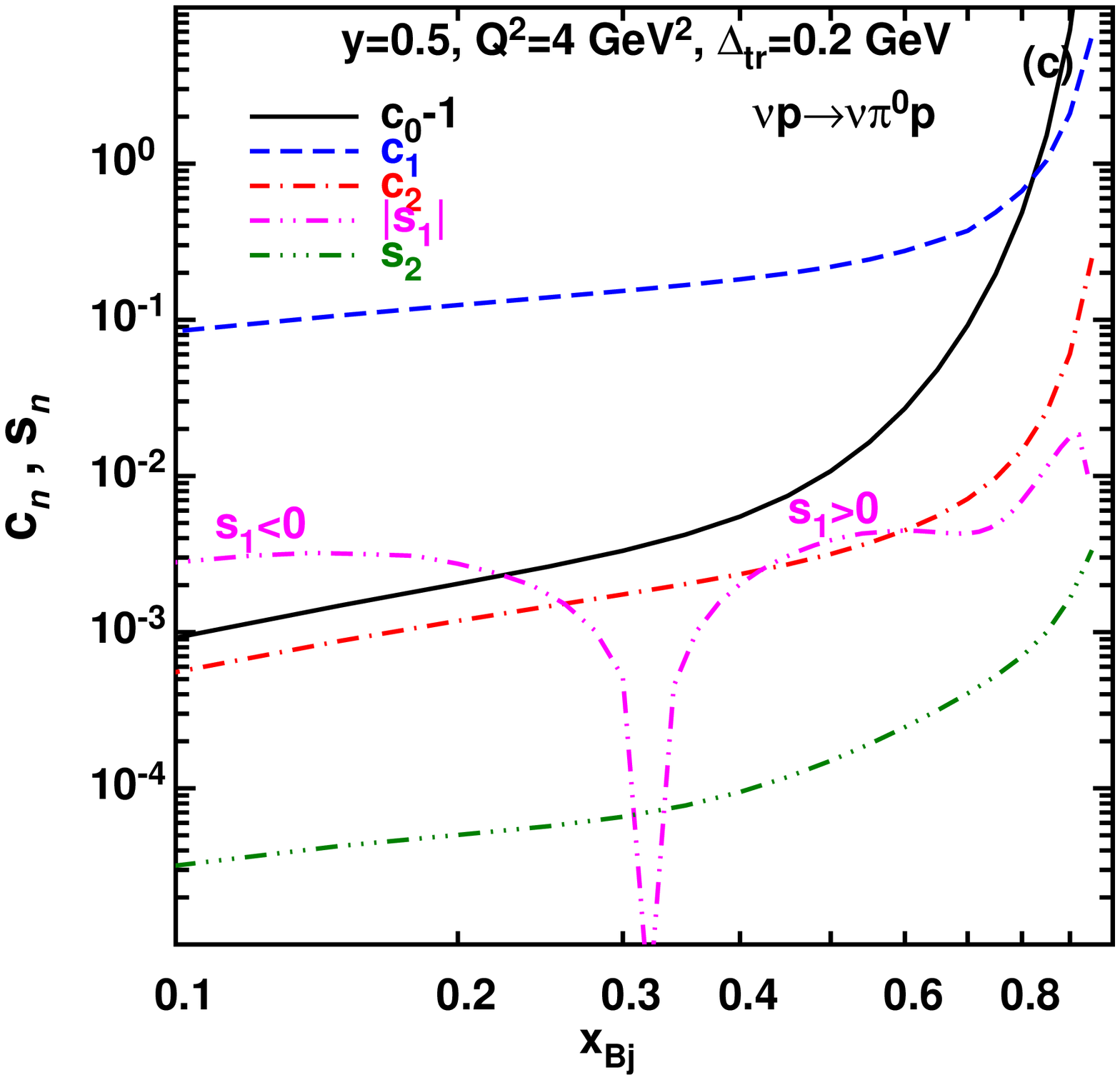}\\
\includegraphics[scale=0.3]{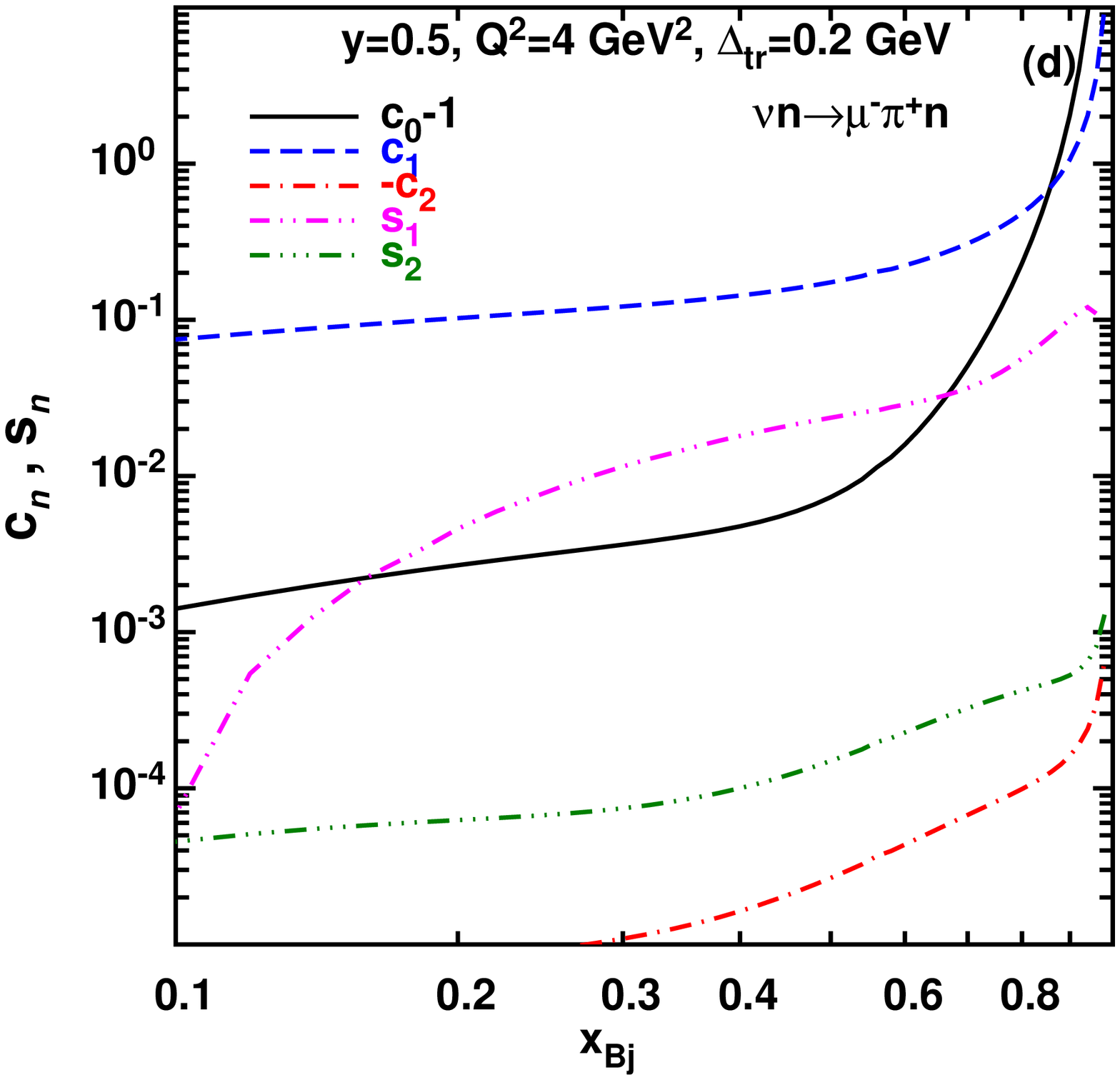}\includegraphics[scale=0.3]{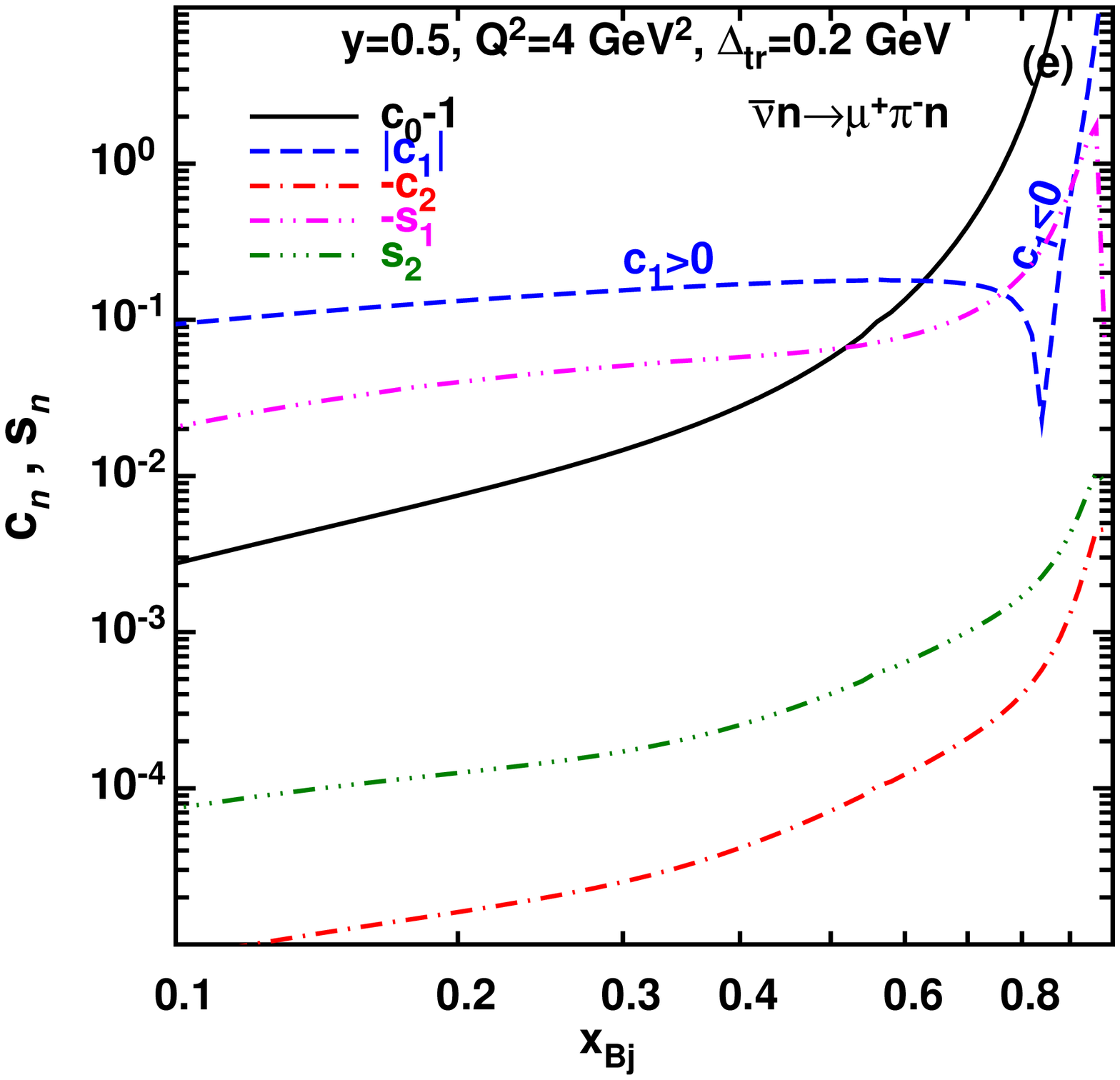}\includegraphics[scale=0.3]{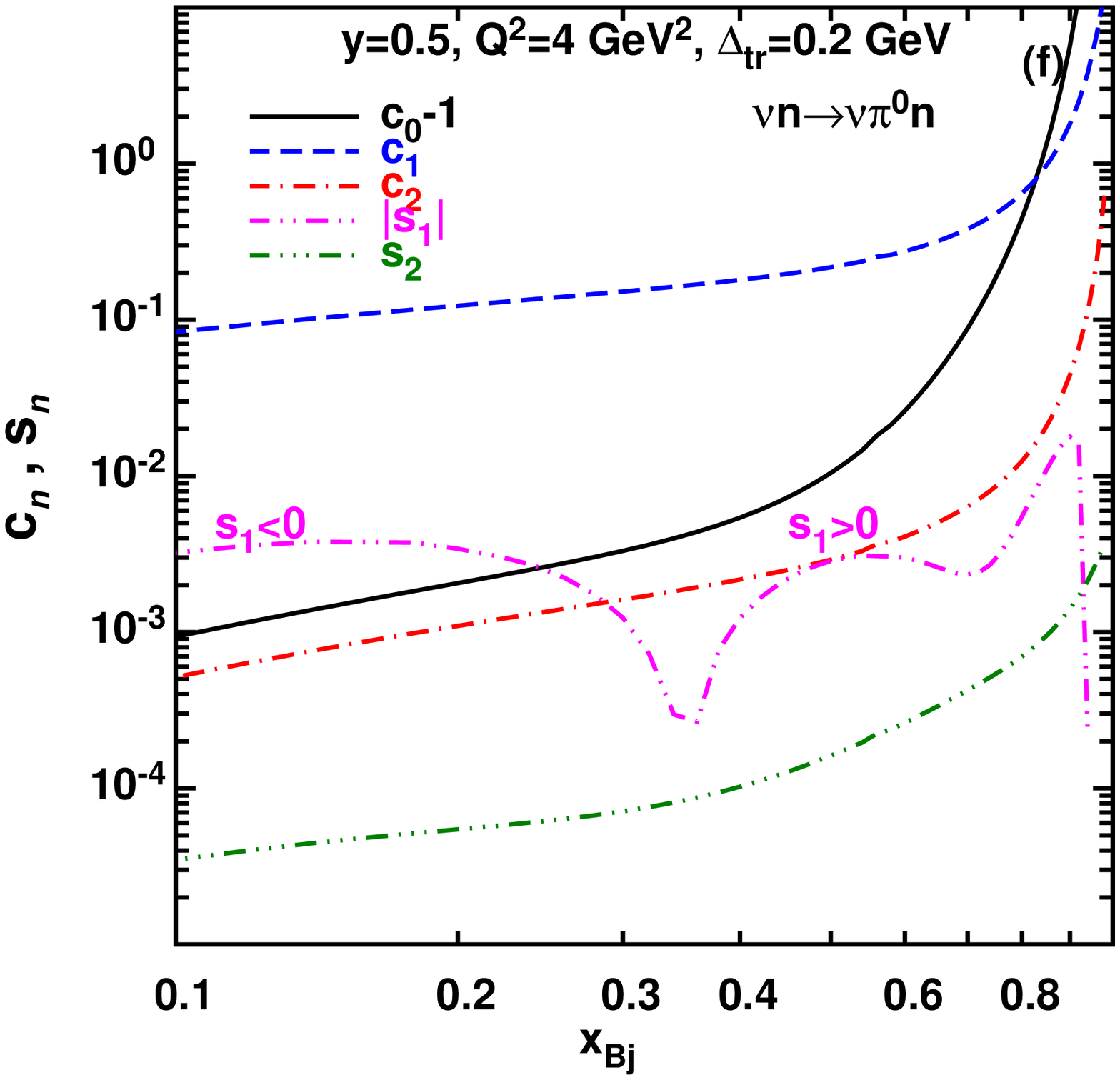}\\

\caption{\label{fig:DVMP-pions}(color online) Pion production on nucleons
without change of the baryon state. Processes in the lower row differ
from the processes in the upper row due to  isospin conservation  breakdown by higher-twist
corrections.}
\end{figure}

In Figure~\ref{fig:DVMP-pions} we show the harmonics $c_{n},\, s_{n}$
for processes without change of baryon state. The processes shown
in the lower row are isospin conjugate to processes in the upper row.
While in the leading twist the cross-sections of the former and the
latter coincide, with the account of the twist-3 corrections this is no longer
valid due to the difference in weak isospin of $\nu$ and $\bar{\nu}$. In
all cases at $x_B\lesssim0.5$, where the cross-section is
the largest, the harmonics are small and do not exceed few per cent.
The largest twist-3 contribution is due to the $c_{1}$ harmonics,
which may reach up to twenty per cent. This is different from the electroproduction
experiments, where $c_{1}$ ($\sim\sigma_{LT}$) is very small. This
result can be understood from~(\ref{eq:sigma_LT}): due to parity
nonconservation in weak interactions we have for the interference
term $\sigma_{0+}\not=\sigma_{0-}$. A positive value of $c_{1}$
for most processes implies that pion production correlates
with the direction of the produced muon (scattered neutrino) in the case of
CC (NC) mediated processes. The interference term also yields a relatively
large harmonics $s_{1}$ which appears due to the interference of the
vector and axial vector contributions. 

In the region of $x\gtrsim0.5$ all the harmonics increase, but the
cross-sections for both the leading twist and subleading twist results
are suppressed there due to increase of $|t_{min}|$ and are hardly
accessible with ongoing and forthcoming experiments.

\begin{figure}
\includegraphics[scale=0.3]{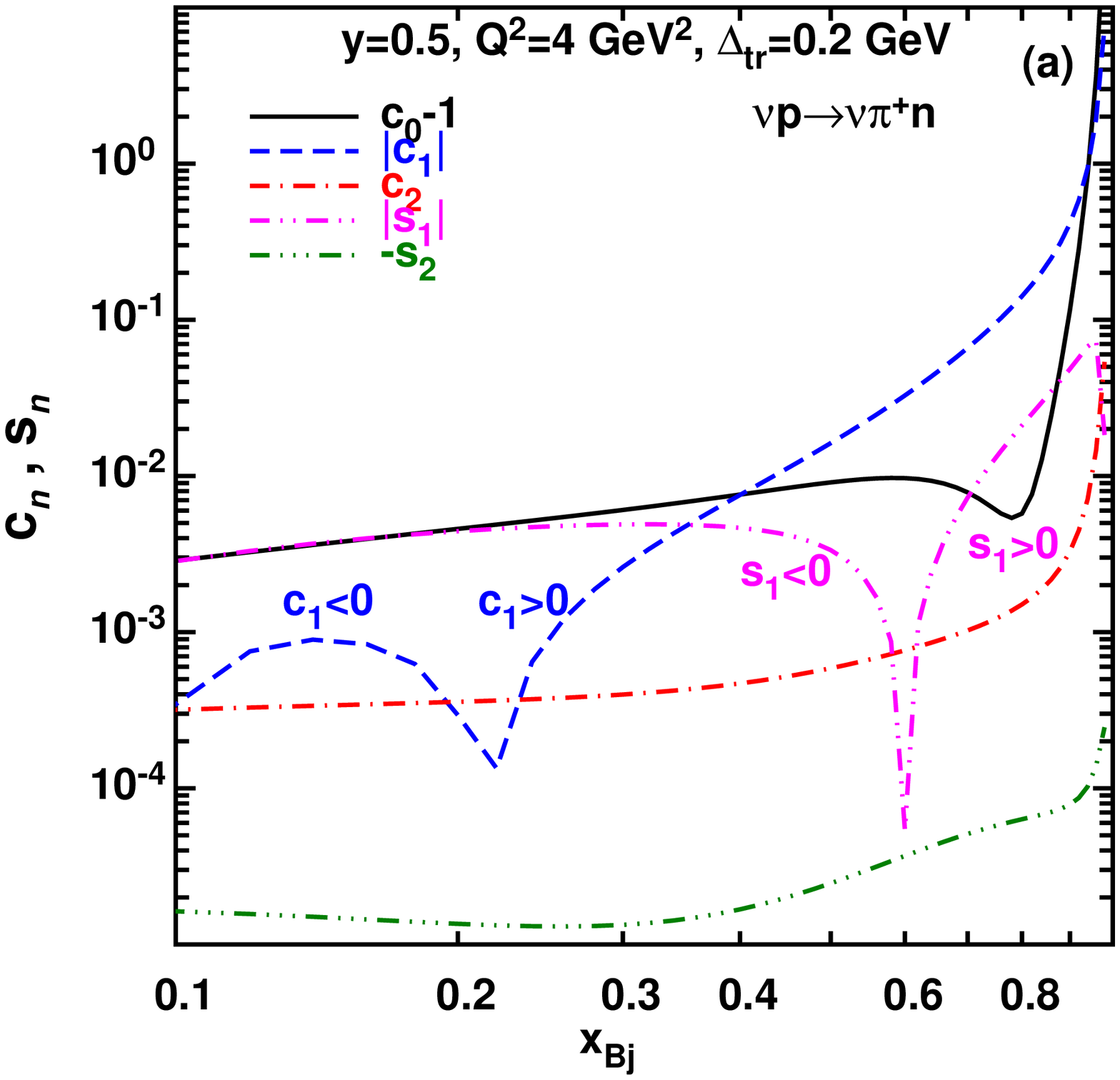}\includegraphics[scale=0.3]{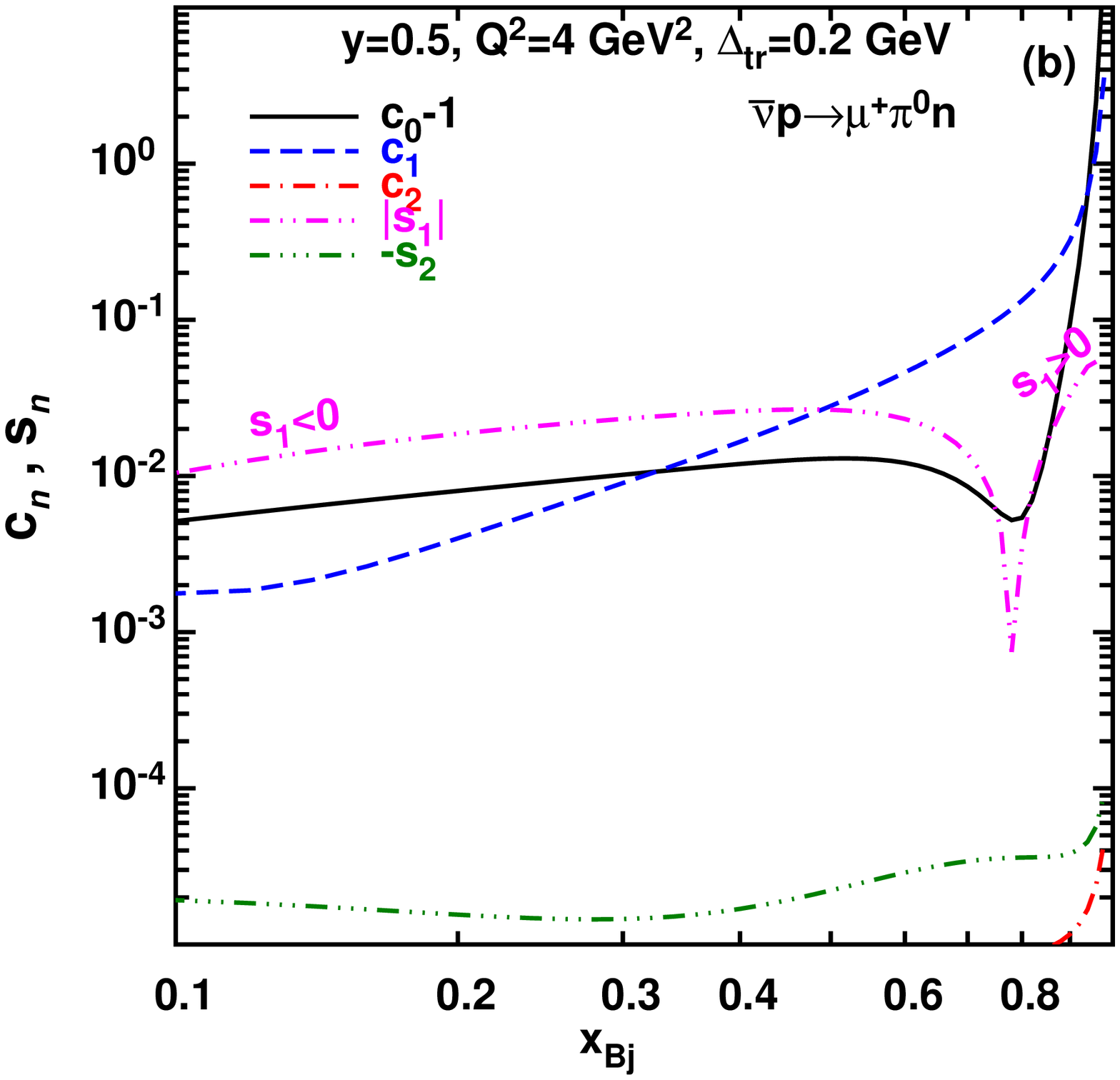}\\
\includegraphics[scale=0.3]{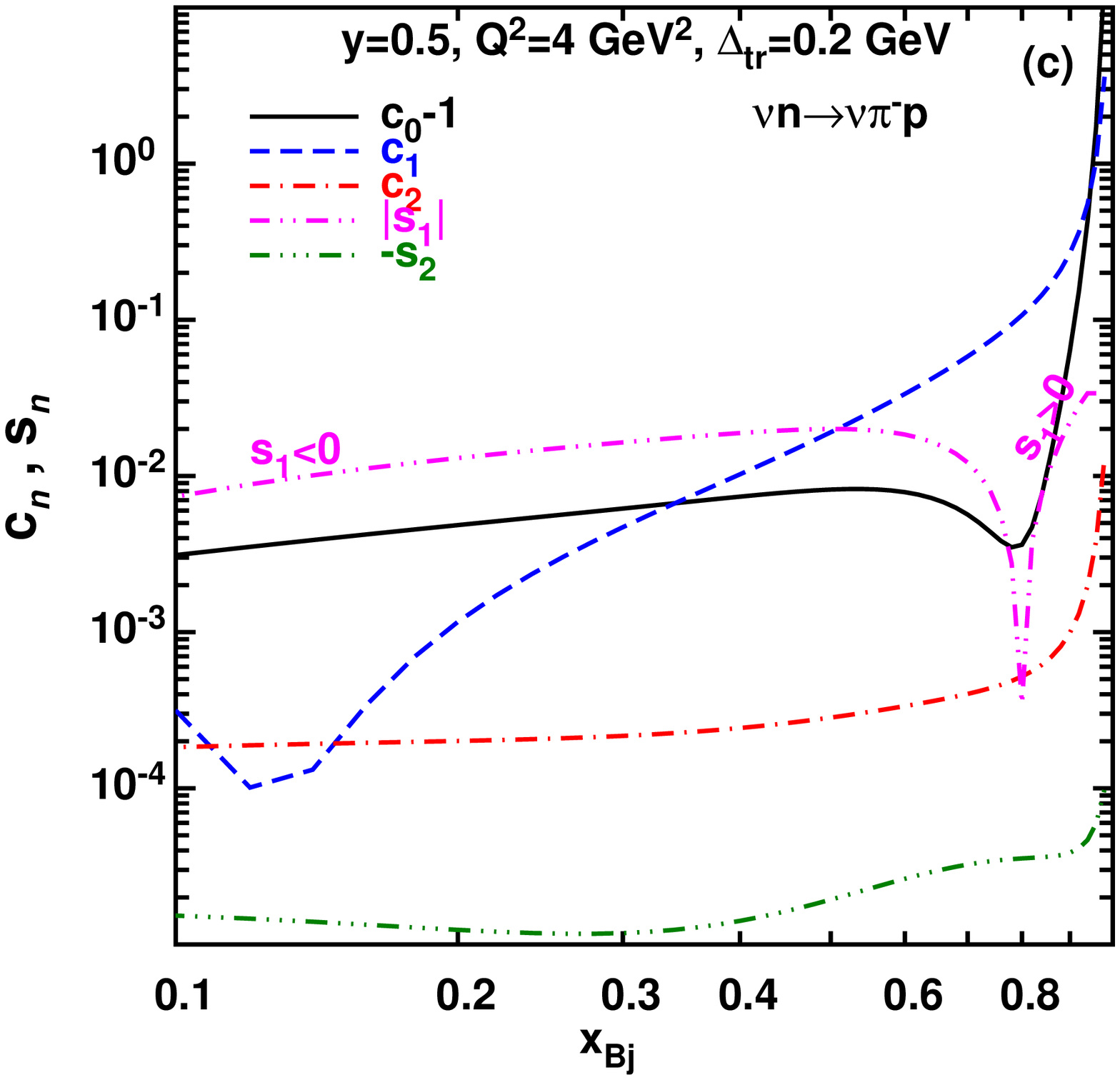}\includegraphics[scale=0.3]{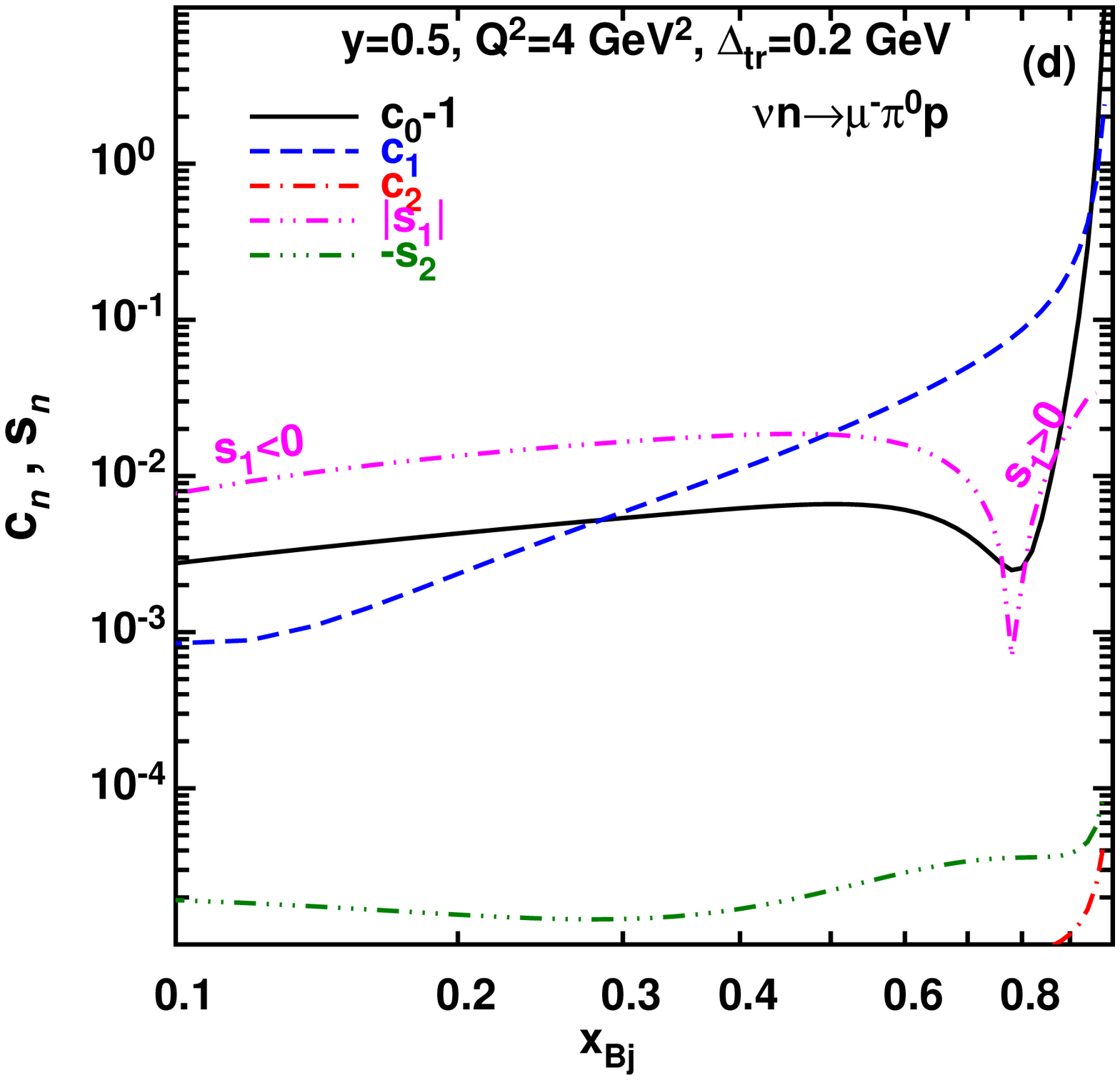}\\

\caption{\label{fig:DVMP-pions-off}(color online) Pion production on nucleons
with change of the baryon state. Processes in the lower row differ
from the processes in the upper row due to isospin-breaking by higher-twist
corrections.}
\end{figure}

In the Figure~\ref{fig:DVMP-pions-off} we present the harmonics
$c_{n},\, s_{n}$ for processes with change of the baryon state. As
was discussed in~\cite{Kopeliovich:2012dr}, in the leading twist
these processes are sensitive to the valence quarks distributions.
As we can see, similar to the previous case, all the harmonics are
small and except the region of $x_{B}\sim1$ do not exceed few per cent.

For strangeness production we obtained qualitatively similar results (corrections
are small), however we refrain from making predictions because
the corresponding amplitudes are sensitive to the strange component
of the chiral odd GPDs which are unknown at this moment.

\section{Conclusions}

In this paper we estimated the contributions of the twist-3 corrections
due to the chiral odd GPDs. One of the manifestations of the twist-3 corrections
is the appearance of the dependence on the angle between the
lepton scattering and pion production planes. We found that the largest
harmonics is $c_{1}$, which can reach up to twenty per cent, however
it does not affect the angular integrated cross-section $d\sigma/dx_{B}\, dt\, dQ^{2}$.
All the other harmonics are small and do not exceed few per cent.
This happens because in case of neutrino interactions, in contrast to
electroproduction of pions, there are large contributions of unpolarized
GPDs $H,\, E$ to the leading-twist amplitude. Notice that the similar
angular harmonics may be generated by interference of the leading
twist result with the electromagnetic corrections~\cite{Kopeliovich:2013ae}.
At moderate virtualities of the order a few ${\rm GeV}^{2}$ this mechanism
also gives small harmonics (of the order few per cent), however those
corrections grow rapidly as a function of $Q^{2}$, and already at
$Q^{2}\sim100\,{\rm GeV^{2}}$ electromagnetic mechanism becomes dominant.

To summarize, we conclude that deeply virtual production of pions and
kaons on protons and neutrons by neutrinos with typical values of $Q^{2}$ of the
order few ${\rm GeV}^{2}$ provide a clean probe for the GPDs, with
various corrections of the order of few per cent. Our results are relevant
for analysis of the pion and kaon production in the \textsc{Minerva
}experiment at FERMILAB as well as for the planned~Muon Collider/Neutrino
Factory~\cite{Gallardo:1996aa,Ankenbrandt:1999as,Alsharoa:2002wu}.
An ideal target for study of the GPDs could be a liquid hydrogen
or deuterium. For other targets there is an additional uncertainty
due to the nuclear effects which will be addressed elsewhere.

We provide a computational code, which can be used for evaluation
of the cross-sections with inclusion of the twist-3 corrections employing
various GPD models.

\section*{Acknowledgments}

This work was supported in part by Fondecyt (Chile) grants No. 1130543,
1100287 and 1120920.

\appendix

 \end{document}